\documentclass[conference]{IEEEtran}
\IEEEoverridecommandlockouts

\usepackage{cite}
\usepackage{amsmath,amssymb,amsfonts}
\usepackage{algorithmic}
\usepackage{graphicx}
\usepackage{textcomp}
\usepackage{xcolor}
\usepackage{booktabs}
\usepackage{longtable}  
\usepackage{multirow} 
\usepackage{float}      
\usepackage{placeins}
\usepackage{ltablex}
\def\BibTeX{{\rm B\kern-.05em{\sc i\kern-.025em b}\kern-.08em
    T\kern-.1667em\lower.7ex\hbox{E}\kern-.125emX}}
\begin{document}

\title{Frontend Diffusion: \\ Empowering Self-Representation of Researchers and Designers with Multi-agent System}

\author{
Zijian Ding$^{1}$,
Qinshi Zhang$^{2}$,
Mohan Chi$^{3}$,
Ziyi Wang$^{1}$ \\
$^{1}$University of Maryland, College Park, 
$^{2}$University of California, San Diego,
$^{3}$Purdue University \\
$^{1}$\texttt{\{ding, zoewang\}@umd.edu},
$^{2}$\texttt{qiz065@ucsd.edu},
$^{3}$\texttt{chi70@purdue.edu}
}

\maketitle

\begin{abstract}
With the continuous development of generative AI's logical reasoning abilities, AI's growing code-generation potential poses challenges for both technical and creative professionals. But how can these advances be directed toward empowering junior researchers and designers who often require additional help to build and express their professional and personal identities? We introduce Frontend Diffusion, a multi-agent coding system transforming user-drawn layouts and textual prompts into refined website code, thereby supporting self-representation goals. A user study with 13 junior researchers and designers shows AI as a human capability enhancer rather than a replacement, and highlights the importance of bidirectional human-AI alignment. We then discuss future work such as leveraging AI for career development and fostering bidirectional human-AI alignment of multi-agent systems.
\end{abstract}



\section{Introduction}
\label{sec:intro}

With the advancements in generative models' logical reasoning capabilities \cite{guo2025deepseek,openaiOpenAIO3mini2025}, technical productivity has been enhanced across various domains. In programming, generative models have increased the resolution rate of pull requests on GitHub from under 2\% in 2023 to over 75\% by mid-2025\footnote{https://www.swebench.com (verified)}. The performance of AI in coding has not only spurred companies' interest in adopting agentic workflows within software development but have also raised concerns among human professionals about the potential for AI to replace their roles. This anxiety has been particularly pronounced in creative industries, where screenwriters face competition from text-generation models \cite{mirowskiCoWritingScreenplaysTheatre2023} and illustrators contend with text-to-image models \cite{koLargescaleTexttoImageGeneration2023,jiangAIArtIts2023,kawakamiImpactGenerativeAI2024}. Such realities motivate us to explore a future where humans and AI coexist synergistically. Instead of automating humans out of the creative process, AI can—and should—act as a catalyst for enhancing self-expression, facilitating more effective personal and professional presentation, and freeing up time for higher-level intellectual pursuits.

This vision holds particular relevance for emerging scholars, such as early-career PhD and master's students, who stand at a critical juncture in their academic journeys. For them, self-presentation is intricately tied to personal growth, skill development, and the formation of academic identities. Despite their wealth of new ideas and scholarly potential, junior researchers often face challenges in building their professional brands and achieving online visibility. Creating a professional website—a platform for presenting one’s research and career aspirations—can be a daunting endeavor, often hindered by technical challenges and time constraints. Multi-agent workflows from design to development and reflection have the potential to address these challenges, not by replacing the researcher's creative authority, but by enabling them to focus on more meaningful pursuits, such as crafting their scholarly narratives and refining their research agendas, ultimately emphasizing their unique academic identities and depth.

To support self-representation through agentic workflows, we developed \textit{Frontend Diffusion}\footnote{Available at: https://github.com/Carolzhangzz/frontendiffusion}, an open-source, end-to-end multi-agent  system transforming user-drawn layouts and thematic prompts into iteratively refined website code. We evaluated this system through a user study involving 13 participants with diverse technical backgrounds. The findings revealed that the AI tool functions not merely as a code generator but as a collaborative partner. Participants emphasized two major themes: AI as a Human Capability Enhancer, Not a Replacement, and Bidirectional Human-AI Alignment.



\begin{figure*}
  \centering
  \includegraphics[width=1\textwidth]{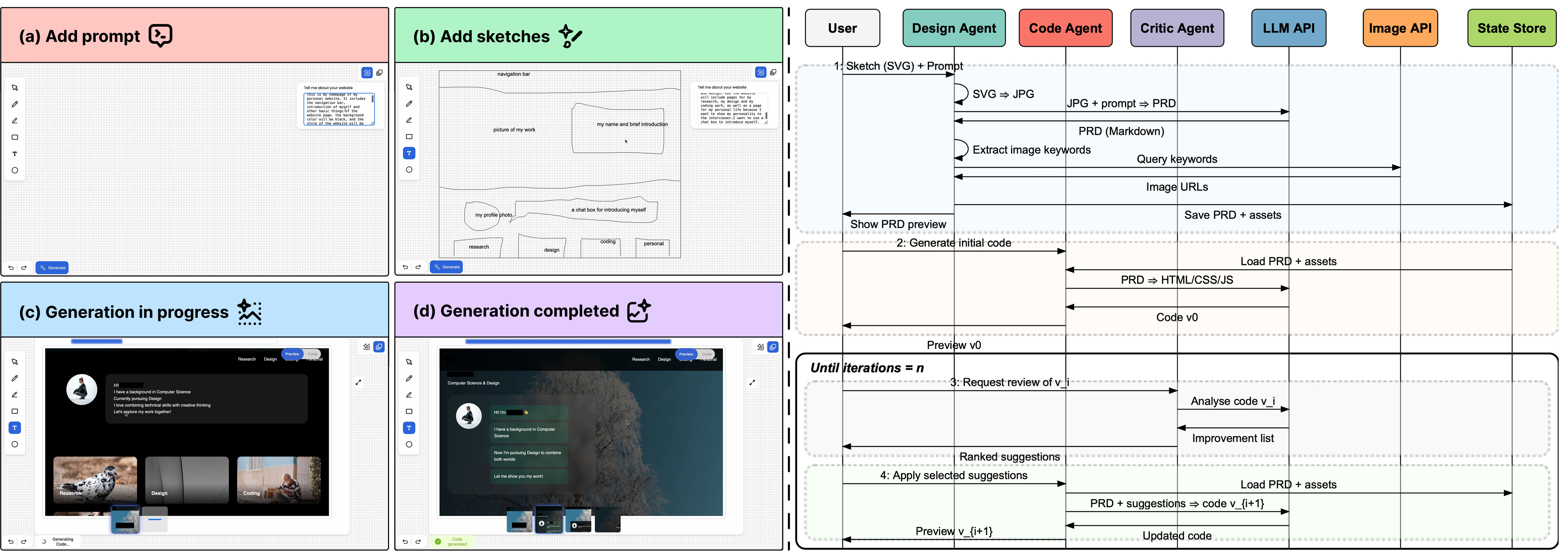}
  \caption{Left: Website generation workflow: (a) user inputs prompt; (b) user draws layout of the website in sketch; (c) the system generates the first website; (d) the system completes generations four website iterations.
  Right: End‑to‑end \textbf{multi‑agent} workflow of \textsc{Frontend Diffusion}. The user supplies a sketch and a textual prompt.
  \textbf{(1) Design Agent} converts the sketch to a Product‑Requirements‑Document (PRD) and retrieves illustrative images;
  \textbf{(2) Code Agent} translates the PRD and assets into runnable HTML/CSS/JavaScript;
  \textbf{(3) Critic Agent} evaluates the code, proposes improvements, and triggers regeneration until the maximum number of iterations $n$ is reached.}
  \label{fig:workflow}
\end{figure*}

\section{Related Work}

The integration of code data in language model training \cite{chenEvaluatingLargeLanguage2021} has made code generation a core capability of generative systems, with front-end code generation demonstrating particularly strong performance \cite{jimenez2024swebench,siDesign2CodeHowFar2024}. Researchers have developed coding agents like OpenHands \cite{openhands} and specialized techniques for UI code generation from screenshots \cite{wanAutomaticallyGeneratingUI2024,wuUICoderFinetuningLarge2024}. Different from traditional template-based approaches, AI-powered UI generation systems now enable personalized and adaptive interfaces. Tools like FrameKit \cite{wu_framekit_2024} and PromptInfuser \cite{petridisPromptInfuserHowTightly2024} support dynamic UI generation, while DesignAID \cite{cai_designaid_2023} and Misty \cite{luMistyUIPrototyping2024} provide conceptual inspiration and remixing capabilities for designers. AI can also offer real-time design feedback \cite{duan_towards_2023,duanGeneratingAutomaticFeedback2024} and automated heuristic evaluations \cite{wuUIClipDatadrivenModel2024,lu_ai_2024}.

Frontend development is beyond code generation and more like creative workflows. In this creative process, AI reduces the time needed for high-fidelity prototyping \cite{edwardsSketch2PrototypeRapidConceptual2024}, lowers experimentation barriers \cite{palaniEvolvingRolesWorkflows2024}, provides inspiration when users have broad but underspecified concepts \cite{rickSupermindIdeatorExploring2023}, and enables parallel prototyping of diverse design solutions \cite{dowParallelPrototypingLeads2010}. Yet, much of the existing literature emphasizes technical prowess over the ways AI can empower particular user groups to shape and communicate their own identities. Building on previous work, our work examines how a multi-agent system can be customized to help junior researchers and designers articulate and curate their professional personas.

\section{System Design}

We developed \textsc{Frontend Diffusion}, an end‑to‑end \emph{multi‑agent} workflow (Figure~\ref{fig:workflow}) that delegates distinct responsibilities to three cooperating agents:


\begin{itemize}
    \item Design Agent (Sketch‑to‑PRD): The Design Agent translates the user’s visual sketch (SVG) and prompt into a structured PRD. After a format conversion of the sketch (SVG $\rightarrow$ JPG) that improves model perception, the agent prompts an LLM to produce a markdown Product Requirements Document (PRD) containing layout, component semantics, and image searching keywords. Keywords in PRD such as [hero(landscape)] or [profile(large)] are extracted and issued to an image searching API\footnote{https://www.pexels.com}, whose URLs are injected back into the PRD. The PRD containing image links are stored in state for downstream agents.
    \item Code Agent (PRD‑to‑Code): Using the stored PRD, image links, and the user original prompt, the Code Agent prompts LLM to generate the initial version of the website ($v_0$).
    \item Critic Agent (Reflection \& Refinement): The Critic Agent performs automated code review and reflection. It analyses the current version ($v_i$), produces improvement suggestions (layout, accessibility, performance, etc.), and iterate on the site ($v_i \rightarrow v_{i+1}$) based on the suggestions. This agentic loop continues for up to $n{=}4$ iterations by default, and users can branch between versions via thumbnail previews.
\end{itemize}

All agents communicate through a shared memory layer and expose JSON‑RPC APIs, making the multi-agent system model‑agnostic and future‑proof against generative models advance. The multi-agent system is based on Claude-3.5-Sonnet (2024‑10‑22), the state-of-the-art language model for coding as of Jan 2025

\section{Study Design}

For evaluating the quality of generated personal websites, besides general objective requirements for websites, subjective factors—such as the content and message the user intends to convey—are more important. Therefore, we used qualitative user interviews to assess both the generated websites and the human-AI co-creation process. The experiment lasted approximately 45 minutes and consisted of three parts: a 5-minute participant onboarding, a 20-30 minute user study, and a 15-20 minute post-study interview. First, participants read and signed the consent form, granting permission for screen recording. Next, the researcher demonstrated how to use it by creating one website page. Participants then created personal website showcasing either professional content, such as research or design portfolios, or personal interests, such as an image gallery or reading list. After completing the user study, the researcher conducted a 15-20 minute post-study interview.

\subsection{Demographic Information}

In the pre-study survey, we collected demographic information, including participants' age, gender, educational level, experience in web development, design, and research, as well as whether they had a personal website (Y/N). Table \ref{tab:demographic} summarizes this information. 


\begin{table}[h!]
  \centering
  \caption{Participants’ demographic information, including age, gender, educational level, and experience in web development (Dev), design, and research (in years).}
  \label{tab:demographic}
  \begin{tabular}{@{}ccccccc@{}}
    \toprule
    & \multicolumn{3}{c}{\textbf{Demographic}}
    & \multicolumn{3}{c}{\textbf{Experience (yrs)}} \\ 
    \cmidrule(lr){2-4} \cmidrule(lr){5-7}
    PID & Age & Gender & Edu & Dev & Design & Research \\
    \midrule
    1  & 25 & F & Yr-1 PhD    & 0   & 0   & 3   \\
    2  & 25 & F & Yr-1 PhD    & 0.1 & 0.2 & 4   \\
    3  & 23 & F & Yr-1 Master & 1   & 3   & 0.5 \\
    4  & 23 & F & Yr-2 Master & 0.5 & 3   & 3   \\
    5  & 28 & F & Yr-3 PhD    & 1.5 & 4   & 5.5 \\
    6  & 31 & M & Yr-3 PhD    & 0.5 & 1   & 10  \\
    7  & 29 & M & Yr-1 PhD    & 0   & 0   & 4   \\
    8  & 23 & F & Yr-1 Master & 1   & 0   & 1   \\
    9  & 31 & F & Yr-4 PhD    & 0   & 1   & 8   \\
    10 & 24 & F & Yr-1 PhD    & 0.5 & 0   & 3   \\
    11 & 23 & F & Yr-1 Master & 3   & 1   & 0   \\
    12 & 29 & M & Yr-3 PhD    & 1   & 0   & 4   \\
    13 & 27 & M & Yr-1 PhD    & 2   & 3   & 3.5 \\
    \bottomrule \end{tabular}
\end{table}


\subsection{Post-Study Interview}
\label{sec:post-study}

Following the completion of the website creation task, we conducted semi-structured interviews with each participant.
The interview outline was as follows:

\begin{itemize}
    \item \textbf{User Experience}
    \begin{itemize}
        \item \textbf{Process:} Did you find the process of using the tool to generate a personal website smooth? Were there any steps that were particularly time-consuming or challenging?
        \item \textbf{Result:} Did the final website meet your expectations? Do you think the tool accurately understood your intent between sketch input and the generated website? Why or why not?
    \end{itemize}

    \item \textbf{Improvement and Recommendation}
    \begin{itemize}
        \item Are there any features you would like the tool to add or improve? Did you feel the need for more interaction or guidance from the tool?
        \item Would you recommend this tool to your peers? Why or why not?
    \end{itemize}

    \item \textbf{Broader Impacts and Perceptions}
    \begin{itemize}
        \item Does this tool increase or decrease your interest in frontend development, or even research?
        \item Do you think this tool enhances human abilities or replaces them? Why?
    \end{itemize}
\end{itemize}

\subsection{Data Analysis}
\label{sec:analysis}

Interview transcripts were analyzed using Braun and Clarke’s \cite{braunUsingThematicAnalysis2006} six-phase thematic analysis framework. First, the data were read repeatedly to gain familiarity and generate initial codes. Codes were then collated into provisional themes, which were reviewed and refined for coherence and distinctiveness. Finally, each theme was defined and illustrated with representative quotations, and analytic memos were kept throughout to ensure transparency and rigor.



\begin{table*}[!htbp]
\caption{Subthemes and representative user quotes for Theme 1 -- AI as a Human Capability Enhancer.}
  \label{tab:results_theme1}
  \centering
  \small
  \begin{tabular}{@{}p{3.5cm}p{14cm}@{}}
    \toprule
    \multicolumn{2}{c}{\bfseries Theme 1 -- AI as a Human Capability Enhancer} \\
    \midrule
    \bfseries Subtheme (Description) & \bfseries Participant Quotes \\
    \midrule
    Website Generation Beyond Templates\par\vspace{0.4em} (AI generates personalized, targeted code)
      & “With templates, you spend time reviewing different options, and even after selection, you must understand their code structure. This tool generates clean, targeted output without unnecessary elements.” (P2)
      \par\vspace{0.4em}
      “For those without UI design experience who struggle with platforms like Wix - which has numerous options requiring drag-and-drop interactions and color decisions - this tool streamlines the process by handling color schemes automatically. When I needed a floating music player, instead of searching through component libraries, the tool generated one directly with multiple variations.” (P7) \\
    \midrule
    Supporting Ideation, Team Brainstorming \& Self-Reflection\par\vspace{0.4em} (AI facilitates collaborative prototyping and self-examination via rapid sketching and text input)
      & 
      “It provides accessibility and facilitates communication. In group projects, team members can individually contribute through text or sketches, collectively evaluate outcomes, and iteratively build upon selected concepts - making it particularly suitable for brainstorming sessions.” (P8)
      \par\vspace{0.4em}
      “From my perspective, a personal blog's essential function lies in its capacity to aggregate and synthesize existing knowledge. This utility isn't necessarily directed toward public consumption - it may serve primarily as a tool for self-examination, illuminating the processes of my internal thought world... The act of presenting to oneself necessarily precedes external presentation to the broader audience.” (P5) \\
    \midrule
    Empowering User Interface Development\par\vspace{0.4em} (Accelerates work for experts and reduces frustration for novices by bridging UI design gaps)
      & “To achieve more specific outcomes, some UI/UX knowledge is essential. For instance, when implementing complex animations like a rotating solar system, rather than using general descriptions, I would specify technical requirements like 'I want one sphere rotating around another sphere.' The AI tool can then better interpret these technical terms like `rotate' to generate the appropriate code.” (P8)
      \par\vspace{0.4em}
      “Frontend development can be very frustrating for some people. It can be very painful. People might literally pull their hands up. So this can certainly be a great stress elimination... I prefer back-end programming. I like writing algorithms... I'm not a big fan of frontend development. And when I was learning front, when I was trying to write frontend code for the first few times, I found it very frustrating.” (P13) \\
    \midrule
    Potential Educational Applications\par\vspace{0.4em} (Acts as a mentor with tutorials/examples, guiding learners toward professional standards)
      & “By studying the code it generates, one can understand how the page is constructed... When I cannot envision certain animations, it provides exemplary solutions that I can learn from regarding how such effects are created.” (P11)
      \par\vspace{0.4em}
      “Because I want to see what recruiters and consultants want... if there's a way that the AI can advise me, not just with the visual, but just with the overall look and feel and content of it. Let's say you're a business looking at my UX research portfolio or description of my career, then is this an OK page for that kind of thing?” (P6) \\
    \midrule
    Potential Research Applications\par\vspace{0.4em} (Expedites prototyping and supports accessible code for specialized groups)
      & “My users are blind users, so I need the frontend to be sufficiently simple for screen readers like Voiceover to work effectively. If it can generate HTML that is both simple and well-structured, it would enhance my productivity.” (P2)
      \par\vspace{0.4em}
      “It proved immensely helpful when we needed to create an audio recording experiment website. Previously, we spent extensive time trying to build it with Wix... In 2020, it took me an exceptionally long time to set it up. This tool makes it much simpler - just designing the interface saved me substantial time, especially for components like the music bar and audio elements.” (P7) \\
    \bottomrule
  \end{tabular}
\end{table*}

\begin{table*}[!htbp]
\caption{Subthemes corresponding to user quotes for Theme 2 -- Bidirectional Human--AI Alignment.}
  \label{tab:results_theme2}
  \centering
  \small
  \begin{tabular}{@{}p{4.5cm}p{13cm}@{}}
    \toprule
    \multicolumn{2}{c}{\bfseries Theme 2 -- Bidirectional Human--AI Alignment} \\
    \midrule
    \bfseries Subtheme (Description) & \bfseries Participant Quotes \\
    \midrule
    AI-initiated: Onboarding Support for New Users\par\vspace{0.4em} (Example sites, demo videos, and tutorials to build proper mental models)
      & “I do think another thing that was challenging was coming up with a certain format for the website, just because I don't have one yet and I don't really know what a website, or what a grad student's website should entail at this point.” (P1)
      \par\vspace{0.4em}
      “Even a tutorial or an example of, like, hey, this is what one person put and this is what - this is all the different images that came out from it. This is an example of what AI can do in this context. I think having that as a reference point for me to look at and say, okay, so the AI can develop - can be this creative even when given this.” (P12) \\
    \midrule
    AI-initiated: Prompt Guidance \& Refinement\par\vspace{0.4em} (Interactive follow-up questions and iterative dialogue instead of one-off outputs)
      & “The little bit of challenge comes with figuring out how to prompt it, because sometimes you need to have some prior experience to know what to expect to minimize the number of iterations. So I didn't know whether I should use certain key terminology, like cards or models or something like that.” (P6)
      \par\vspace{0.4em}
      “I definitely think that it should ask maybe follow-up questions when generating the response...Follow up questions to maybe prompt the user to specify what they want.” (P1) \\
    \midrule
    User-initiated: Fine-Grained Control Over Details\par\vspace{0.4em} (Detailed element specs, hierarchical prompts, and previews e.g. colors, images.)
      & “I wonder if you can have a general prompt and then prompt also in the specific section with a little tag.” (P6)
      \par\vspace{0.4em}
      “Some users may be hesitant to share their materials directly with AI systems. They might prefer generating just the structure, then populating it with their information locally... You could have a large model generate an overall template that's sophisticated and well-structured, clearly indicating where user information should go, then use a local small model to input the personal content without uploading sensitive materials to servers.” (P2) \\
    \midrule
    User-initiated: Harnessing AI’s Unexpected Creative Sparks\par\vspace{0.4em} (Captures serendipitous ideas and enables remix/editing across versions)
      & “The AI's knowledge base may exceed humans', potentially offering unexpected ideas...It introduces dynamic effects that I hadn't considered when creating static sketches.” (P11)
      \par\vspace{0.4em}
      “With three or four versions, I might appreciate certain design elements from each. How could we combine them? Perhaps through editing capabilities, comments, or drag-and-drop functionality?” (P4) \\
    \midrule
     User-initiated: Communicating Dynamic Behaviors\par\vspace{0.4em} (Challenges expressing interactions via static sketches/text; needs flowcharts or state diagrams)
      & “While my color instructions were minimal, the system comprehended that I wanted vibrant colors and a clean design. Although my sketch was static, it understood which elements should be dynamic, such as the floating music player that follows page scrolling - an option I hadn't explicitly requested but found valuable.” (P7)
      \par\vspace{0.4em}
      “The underlines here aren't actual content but rather elements meant to be toggled - expressing such toggle requirements in sketches poses a challenge.” (P12) \\
    \bottomrule
  \end{tabular}
\end{table*}

\section{Results}
\label{sec:results}
  
We identified two major themes in our results: Theme 1 - AI as A Human Capability Enhancer and Bidirectional Human--AI Alignment.

Theme 1: When supporting the self-representation of junior researchers and designers, AI has the potential to enhance human capabilities by alleviating the cognitive and technical burden of repetitive tasks. For example,  P3 addressed: ``Without this tool, maybe very few people would have their own website; but once it's available, everyone might want one as a digital business card... With such tools, everyone can become a designer, frontend developer, or product manager, thereby driving the creation of more interesting products." This enables users to dedicate more time to higher-level ideation and reflection: ``it's actually picking up intention... it's not going to try to second guess me" (P9), while P10 emphasized the irreplaceable role of humans as requirement providers, noting that ``even the user doesn't have a standard perspective. The users are like the client, and this tool is like the contractor." Detailed supporting quotes for each subtheme can be found in Table \ref{tab:results_theme1}.

Theme 2 (Table \ref{tab:results_theme2}): We decompose this into subthemes based on the initiator of the alignment process as below:

\begin{enumerate}
    \item \textbf{Ai-initiated alignment}
    \begin{itemize}
        \item Onboarding support for new users.
        \item Prompt guidance and refinement.
    \end{itemize}
    
    \item \textbf{User-initiated alignment}
    \begin{itemize}
        \item Fine-grained user control over details.
        \item Harnessing AI's unexpected creative sparks.
        \item Communicating dynamic behaviors.
    \end{itemize}
\end{enumerate}

Notably, participants (P2, P6, P8, P10, P13) emphasized the need for fine‐grained user control over details. Although the multi‐agent system is developed to interpret high‐level sketches and prompts for end‐to‐end website generation, users expressed a need for more granular control over intent expression. For example, P2 wanted to provide more precise parameters or specific regions. Similarly, P6 advocated for hierarchical prompting for different page sections: ``I wonder if you can have a general prompt and then prompt also in the specific section with a little tag." More detailed participant quotes for each subtheme are provided in Table \ref{tab:results_theme2}.

\section{Discussion}




\subsection{AI as a Human Capability Enhancer: AI Career Advising and Planning}

The results demonstrate that generative AI systems can actively enhance users’ self-presentation and career planning by streamlining technical burdens and simulating industry perspectives. P6 explicitly noted that the AI tool could role-play recruiter and consultant viewpoints—offering concrete guidance on overall look, feel, and content that aligns with professional expectations. By providing a low-cost “try-on” environment for career scenarios, AI reduces both cognitive load and technical barriers, thereby empowering users to experiment with diverse pathways.

\subsection{Bidirectional Human–AI Alignment: Alignment in Hierarchical Multi-agent Systems}

As AI reasoning and planning capabilities advance and multi-agent systems scale, they can decompose complex human objectives into extended task sequences and achieve more significant outcomes. In our interviews, participants underscored the importance of onboarding support for novice users, guidance and refinement of prompts, and fine-grained control over agent behavior. Enabling precise user intervention—such as hierarchical prompting \cite{gmeinerIntentTaggingExploring2025}, where a general instruction is augmented by section-specific tags—requires a transparent multi-agent architecture: in our system, each agent’s role (Design, Code, or Critic) should be explicitly defined and visible to users. Moreover, the interface should allow users to address individual agents directly—for example, requesting additional design variations from the Design Agent or further review iterations from the Critic Agent—thereby facilitating targeted interventions and reinforcing bidirectional alignment by interacting with each agent.  A promising avenue is to explore coordination both between humans and agents and among agents themselves \cite{baiMASMARSCoordination2025}.

\subsection{Limitation and Future Work}

There were several limitations in the current study that warrant further investigation. First, the user sample was relatively narrow with 13 junior researchers and designers. This sample did not encompass a broader demographic, such as participants from non-academic backgrounds. Second, this study leveraged publicly available online image repositories when generating website content to avoid direct use of participants' personal images. This approach resulted in the selection of images that participants perceived as irrelevant or misaligned with their personal or professional identity, potentially impacting the overall user experience. Future work could not only expand the sample size and enhance population diversity but also pursue the following three research directions:
\begin{itemize}
  \item Studying how AI systems could augment—rather than replace—human capabilities over extended periods, for example in supporting career planning and skill development;
  \item Increasing the transparency of multi-agent systems to enable users to employ hierarchical prompting strategies to each individual agent;
  \item Developing localized solutions—such as open‐weight models—to allow sensitive personal data to be processed and stored on users’ local devices.
\end{itemize}
In conclusion, we eagerly look forward to further research exploring the real-world impact of multi-agent systems.




\newpage

\bibliographystyle{IEEEtran}
\bibliography{reference}

\appendices

\end{document}